\documentclass[prl,twocolumn,showpacs,floatfix,preprintnumbers,amsmath, amssymb,superscriptaddress]{revtex4}

\usepackage{graphicx}
\usepackage{amsmath}
\usepackage{amssymb}
\usepackage{color}
\usepackage{dcolumn}
\usepackage{epsfig}
\usepackage{bm}
\usepackage[urlcolor=blue]{hyperref}
\hypersetup{backref, colorlinks=true, linkcolor=blue, citecolor=blue}

\begin{document}

\title{Superconductivity in Solid Benzene Molecular Crystal}

\author{Guo-Hua Zhong}
\affiliation{Shenzhen Institutes of Advanced Technology, Chinese Academy of Sciences, Shenzhen 518055, China}
\affiliation{Beijing Computational Science Research Center, Beijing 100193, China}

\author{Chun-Lei Yang}
\affiliation{Shenzhen Institutes of Advanced Technology, Chinese Academy of Sciences, Shenzhen 518055, China}

\author{Xiao-Jia Chen}
\email{xjchen@hpstar.ac.cn}
\affiliation{Center for High Pressure Science and Technology Advanced Research, Shanghai 201203, China}

\author{Hai-Qing Lin}
\email{haiqing0@csrc.ac.cn}
\affiliation{Beijing Computational Science Research Center, Beijing 100193, China}

\date{\today}

\begin{abstract}
Light-element compounds hold great promise of high critical temperature superconductivity judging from the theoretical perspective. Hydrogen-rich material, benzene, is such a kind of candidate but also an organic compound. A series of first-principles calculations are performed on the electronic structures, dynamics properties, and electron-phonon interactions of solid benzene at high pressures. Benzene is found to be dynamically stable in the pressure range of $180-200$ GPa and to exhibit superconductivity with a maximum transition temperature of 20 K at 195 GPa. The phonon modes of carbon atoms are identified to mainly contribute to the electron-phonon interactions driving this superconductivity. The predicted superconductivity in this simplest pristine hydrocarbon shows a common feature in aromatic hydrocarbons and also makes it a bridge to organic and hydrogen-rich superconductors.
\end{abstract}


\maketitle

Exploring new materials with high critical temperature ($T_c$) superconductivity is one attractive field in modern condensed matter physics. Although there has been no specific method to guide the search for superconductors, the light elements or their bearing compounds such as hydrogen and hydrogen-rich materials are believed to be promising candidates. The basis of the latter is mainly from Ashcroft's theoretical viewpoint that hydrogen dominant metallic alloys can reduce the metallization pressure to the scope of the experiment and become the potential candidates of high-$T_{c}$ superconductors \cite{ref1}. In 2008, Chen \emph{et al.} \cite{ref2} observed the signature of metallization of solid silane (SiH$_4$) at pressure above 60 GPa. Subsequently, Eremets \emph{et al.} \cite{ref3} measured superconductivity with $T_c\sim 17$ K in solid SiH$_4$, though there are still some queries about these findings. These experimental observations imply the feasibility of Ashcroft's theory. Recently, superconductivity at around $T_c\sim 200$ K was experimentally observed in the sulfur hydride system \cite{ref4,ref5}, illustrating the feasibility of seeking for the high-$T_{c}$ superconductors in hydrogen-rich materials. Prior to this discovery, Duan \emph{et al.} \cite{ref6} predicted the H$_3$S ($Im$-$3m$) structure for sulfur hydride at high pressure. These authors predicted potential superconductivity with high $T_c$ value of $191-204$ K at 200 GPa \cite{ref6}. The predicted crystal structure was soon confirmed in experiments \cite{ref7,ref8,ref9}, though many other sulfur hydrides form such as H$_4$S$_3$, H$_5$S$_8$, H$_3$S$_5$, and HS$_2$. It has also been suggested \cite{ref6} that the electron-phonon coupling mainly arises from H vibrations in this hydride. For instant, the contribution of H vibrations to the coupling reaches to 90\% in the sulfur hydride system. These results demonstrate the extreme importance that pressure played in the discovery of superconductivity of light materials as well as in hybridizing interaction of H with other elements.

In addition, organic based compounds were also suggested as candidates of high temperature or room temperature superconductors \cite {ref10,ref11}. This idea assumed that the interaction of electrons with much higher excitation energy than the phonon energy can result in a substantially higher $T_{c}$. The first experimental evidence of superconductivity in organic metals was found in 1980 \cite{ref12}. Since then, numerous organic superconductors have been reported including electron donor and electron acceptor molecules. Superconductivity with $T_c$ as high as 38 K was observed in cesium-doped fullerene \cite{ref13}, and more than 120 K was also found in potassium-doped $p$-terphenyl \cite{ref14}. These findings highlight that organic compounds have potential to become high-$T_c$ superconductors. This is mainly because non phonon mechanisms were found to account for their superconductivity in these kinds of materials. Only electron-phonon coupling is not enough to produce such high-$T_c$ superconductivity in alkali metal doped fullerides \cite{ref13} and PAHs \cite{ref15}. Meanwhile, the low dimensional feature of organic molecule was proposed to favor the strong electronic correlation effects in these materials such as doped fullerides \cite{ref16} and PAHs \cite{ref17,ref18,ref19}. While pressure can greatly enhance superconductivity in PAHs such as potassium-doped phenanthrene \cite{ref20} and picene \cite{ref21}, the realization of superconductivity solely by applying pressure on pristine organic compounds has not been achieved yet.

In this work, we choose benzene (C$_6$H$_6$), a hybrid of the simplest aromatic hydrocarbon and a hydrogen-rich material, to explore superconductivity at high pressures. Solid C$_6$H$_6$ was previously predicted to enter a metallic state at pressures above $180$ GPa \cite{ref22}, though the metallization had not been realized experimentally \cite{ref23}. Through extensive calculations of electronic structures, dynamical properties and electron-phonon interactions, we find that C$_6$H$_6$ is in fact superconducting in the pressure range of 180 and 200 GPa with a maximum $T_{c}$ of 20 K. Within the framework of electron-phonon coupling, the superconductivity is examined to mainly come from the contribution of C element, differing from the H dominant materials. Based on our systematical investigations on the superconductivity of potassium-doped PAHs in recent years\cite{ref24}, we conclude that the materials containing benzene rings must be superconducting, with $T_c\sim5-7$ K, and the superconductivity is increased with the change of structure, electronic correlations and pressure.

To study the structural and electronic properties of solid C$_6$H$_6$, we employed the Vienna ab initio simulation package (VASP) \cite{ref25} based on the projector augmented wave method \cite{ref26}. For the plane-wave basis-set expansion, an energy cutoff of 800 eV was adopted. Dense \emph{k}-point meshes were used to sample the first Brillouin zone and ensured that energies converged to within 1 meV/atom. At the same time, lattice dynamics and electron-phonon interactions were calculated using density functional perturbation theory \cite{ref27} and the Troullier-Martins norm-conserving potentials \cite{ref28}, as implemented in the QUANTUM-ESPRESSO (QE) code \cite{ref29}. The cutoff energies of 80 and 600 Ry were used for wave functions and charge densities, respectively. $24\times24\times24$ Monkhorst-Pack \emph{k}-point grid with Gaussian smearing of 0.003 Ry was used for the electron-phonon interaction matrix element calculation at $6\times6\times6$ \emph{q}-point mesh. In both VASP and QE codes, the local density approximation (LDA) \cite {ref30,ref31} functional was selected. Forces and stresses for the converged structures were optimized and checked to be within the error allowance of the VASP and QE codes. The computational methods have been proved to be reliable in previous reports \cite {ref6,ref22}.

\begin{figure}
\includegraphics[width=\columnwidth]{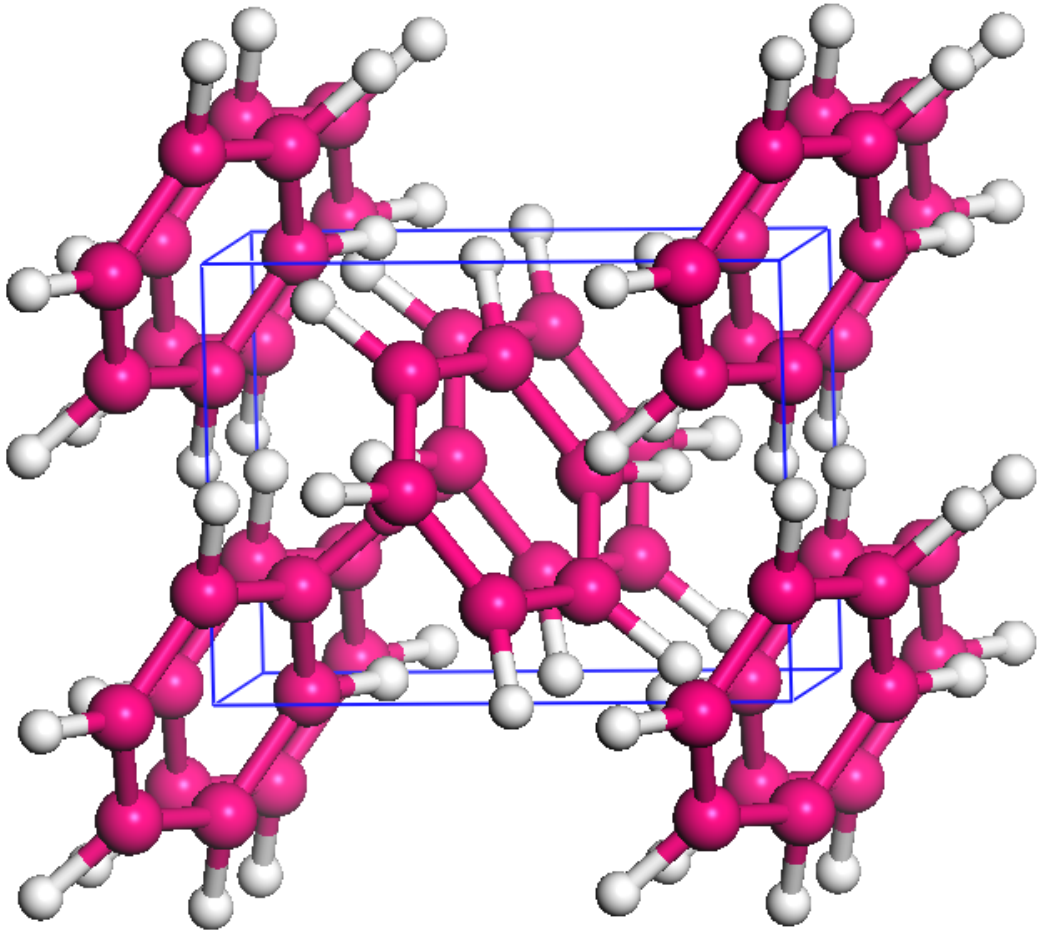}
\caption{(Color online) The geometrical structure of C$_6$H$_6$ crystal in the high pressure range of $180-200$ GPa. Bed and white spheres are represent C and H atoms, respectively.}
\end{figure}

For solid C$_6$H$_6$, it has well-known that the pressure can change its morphology and structure and lead to a series of phase transitions \cite{ref32,ref33,ref34}. At 1.4 GPa, the transition occurs from $Pbca$ phase to $P4_32_12$ phase. At 4 GPa, phase II transfers to $P2_1/c$ phase, and up to 11 GPa. However, at higher pressure above 11 GPa, the experimental results are still not perfect. Keeping the molecular characteristic of benzene, the previous theoretical study pointed out that \cite{ref22}, solid benzene transforms to $P2_1$ from $P2_1/c$ at 40 GPa and remains to 300 GPa. Noticeably, from the energy point of view, the graphane-like crystal possesses the lower enthalpy than molecular benzene crystal starting from about 10 GPa to 300 GPa. $P2_1$ and $P2_1/c$ are only two metastable phases in the high pressure region \cite{ref22}. However, it is not easy to convert benzene to polymer or amorphization compound. Because there are likely significant barriers among these interconverting processes and the conversion reaction also involves temperature condition except for pressure \cite{ref22,ref35}. Thus, both $P2_1$ and $P2_1/c$ phases can keep the feature of benzene molecule instead of the amorphous structure of C and H at high pressure. Especially, $P2_1/c$ phase of C$_6$H$_6$ can behave as a metal in the pressure range of $180-200$ GPa \cite{ref22}.

\begin{figure}
\includegraphics[width=\columnwidth]{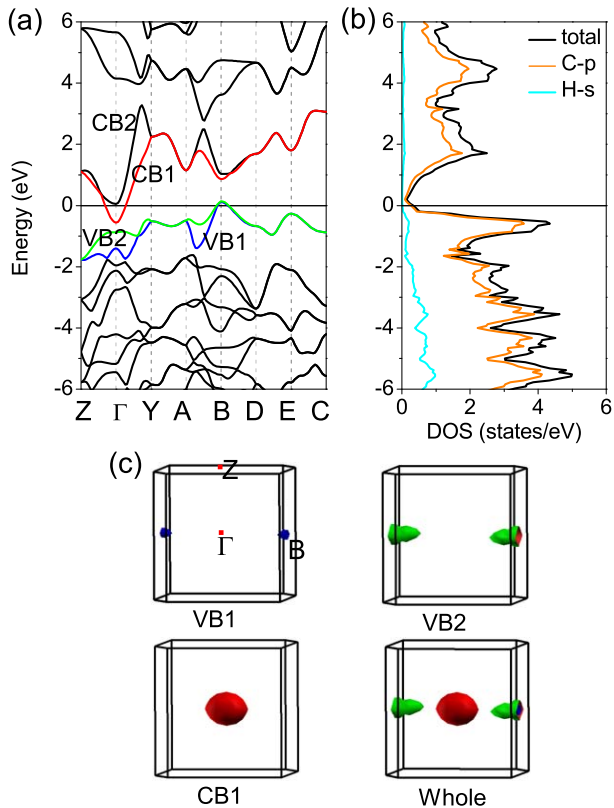}
\caption{(Color online) Electronic structures of C$_6$H$_6$ at 190 GPa: (a) energy band structure, (b) total and projected density of states (DOS) and (c) Fermi surfaces. Zero energy denotes the Fermi level.}
\end{figure}
As a comparison and a check on accuracy, we start our investigation by looking at the first solid phase of benzene with $Pbca$ symmetry. Our optimized crystal constants are respectively $a=7.041$ {\AA}, $b=8.903$ {\AA} and $c=6.357$ {\AA} at zero temperature, which are 3-6\% less than experimental values at 78 K \cite{ref36}. If the effect of temperature is considered, the error between theoretical prediction and experimental measurement is acceptable. Moreover, for $Pbca$ C$_6$H$_6$, our calculated band gap of 4.1 eV is in a good agreement with previous results \cite{ref22}. Based on the test, we have optimized the crystal structures of $P2_1/c$ C$_6$H$_6$ in the range of $180-200$ GPa. Figure 1 shows the geometrical structure of $P2_1/c$ phase of C$_6$H$_6$ crystal. In the case of 190 GPa, the optimized lattice constants $a=3.962$ {\AA}, $b=3.881$ {\AA} and $c=5.287$ {\AA} as well as the angle $\beta=100.3^{\circ}$.

The calculated band gap of 4.1 eV for C$_6$H$_6$ is far less than 7.5 eV of methane (CH$_4$) \cite {ref37}, which indicates C$_6$H$_6$ is more easily to become into metal under pressure comparing with CH$_4$. Indeed, the band gap of C$_6$H$_6$ with $P2_1/c$ structure has been closured when the pressure increases to 180 GPa, and keeping the metallic behavior up to 200 GPa. On the contrary, until 520 GPa, CH$_4$ is still a semiconductor \cite{ref38}. In the case of 190 GPa, we show the electronic band structures, density of states (DOS) and Fermi surface sheets of C$_6$H$_6$ in Fig. 2. As shown in Fig. 2(a), similar to other hydrogen-rich materials, the metallization of C$_6$H$_6$ are mainly derived from the increase of covalent interaction under pressure. However, the nature of energy bands of C$_6$H$_6$ is different from those of doped PAHs where the band structure possessed the typical charge transfer characteristics. Corresponding to band structure, a small amount of electronic states gathers at Fermi level in the DOS picture, which is mainly contributed by C-2\emph{p} states [Fig. 2(b)]. The DOS at Fermi level is about 0.32 states/eV, which is a small value comparing with those of doped PAHs. But, we find that the DOS at Fermi level continuously increases with the increase of pressure in this range of $180-200$ GPa. Checking the fine electronic feature near Fermi level, we find additionally that the VB1 and VB2 bands [marked in Fig. 2(a)] form the hole-like Fermi surfaces around $B$ \emph{k}-point, as shown in Fig. 2(c). The CB1 band with higher energy forms the electron-like Fermi surface around $\Gamma$ \emph{k}-point.

\begin{figure}
\includegraphics[width=\columnwidth]{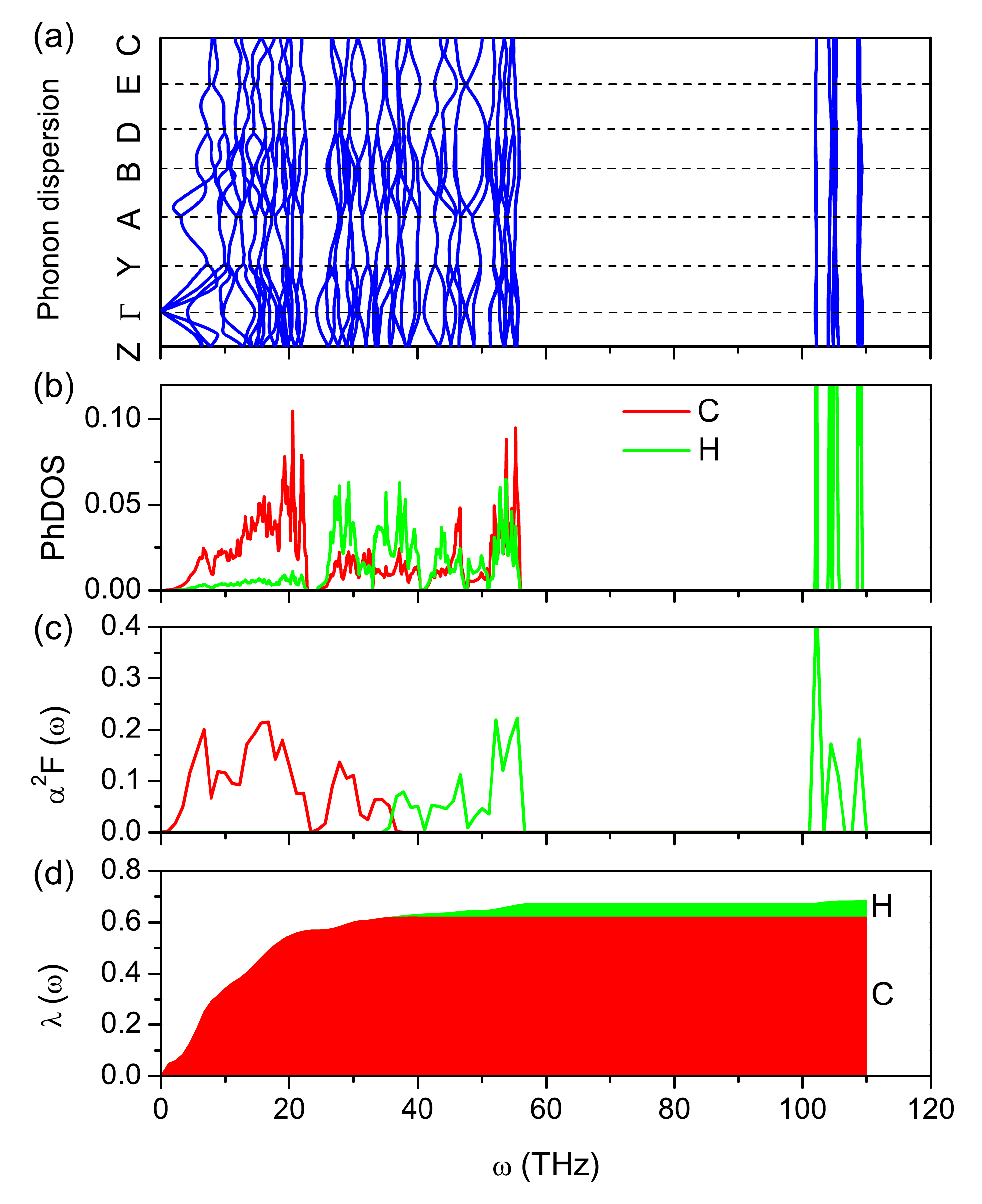}
\caption{(Color online) Calculated phonon dispersion curves (a), phonon density of states (b), Eliashberg spectral function $\alpha^2F(\omega)$ (c) and electron-phonon coupling integral $\lambda(\omega)$ resolved by projections on C and H phonon (d) for $P2_1/c$ structure of benzene at 190 GPa.}
\end{figure}

For this metallic phase of C$_6$H$_6$, our calculated phonon dispersion [Fig. 3(a)] has confirmed the dynamical stability of $P2_1/c$ structure due to the absence of imaginary frequency modes. The visible gaps in the phonon spectra divide phonon frequency into three parts, low-frequency below 22.7 THz, intermediate-frequency from 24.3 to 56.0 THz and high-frequency above 102.0 THz. Combining with the the projected phonon density of states (PhDOS) shown in Fig. 3(b), we can determine that low-frequency modes come mainly from the C atomic motions, the intermediate-frequency bands are induced by the organic intermolecular coupling, while the high-frequency spectrum belongs to the C-H bond stretching vibrations. Corresponding to the origin of phonon modes, the calculated results show that the higher pressure leads to the slightly hardening of intermediate/high-frequency phonon and the softening of low-frequency phonon.

Further understanding the dynamical properties of C$_6$H$_6$ under pressure, we compared it with typical hydrogen-rich materials and organic compounds. For SiH$_4$, the material exists in form of multifold-coordinated silicon hydride at 200 GPa \cite{ref39}. The low-frequency induced by the motions of Si is below 19 THz, the intermediate-frequency region is between 19 and 33 THz, dominated by intermolecular (Si-H-Si) interactions, while the high-frequency phonon from Si-H stretching vibrations is in the range of $33-75$ THz. This implies stronger stretching vibration of C-H bond in C$_6$H$_6$ than Si-H bond in SiH$_4$. Between C$_6$H$_6$ and SiH$_4$, their regions of low-frequency from the heavier element are almost accordant, but there is big difference for the intermediate- and high-frequency modes. For germane (GeH$_4$), the multifold-coordinated germanium hydride and the quasi H$_2$ molecules coexist in the system at 220 GPa \cite{ref40}. The phonon frequency of GeH$_4$ is also obviously divided into three parts similar to C$_6$H$_6$, due to the existence of gaps in phonon energy. The vibration modes of heavier element Ge is below 14 THz, the intermediate-frequency region is between 14 and 62 THz owing to the intermolecular H$_2$ coupling and the Ge-H stretching vibrations, while the high-frequency phonon from the intramolecular H$_2$ vibrations is in the region of $73-80$ THz. Comparing with GeH$_4$, both phonon vibrations and intermolecular coupling in C$_6$H$_6$ are stronger. In H$_2$-containing compounds such as PbH$_4$(H$_2$)$_2$, the alloy like compound of Pb and quasi H$_2$ units is formed at 200 GPa \cite{ref41}. The low-frequency phonon below 8.6 THz comes from heavier element Pb. But the region of intermediate-frequency vibrations in $8.6-56$ THz arises from the intermolecular H$_2$ coupling. Two regions of high-frequency in ranges of $81-87$ THz and $96-101$ THz are induced by the intramolecular vibrations of two kinds of quasi H$_2$ molecules. Comparing with PbH$_4$(H$_2$)$_2$, C$_6$H$_6$ has almost the same intermediate-frequency phonon coupling, but stronger low-frequency vibrations. For sulfur hydride, H$_3$S is the main form at 200 GPa with $Im\bar{3}m$ space-group \cite{ref6}. At this pressure, the vibration modes below 18 THz are due to the motions of S atom, the other phonon bands between 18 and 55 THz are formed by the S-H bond stretching vibrations. There is no higher phonon frequency than 55 THz in H$_3$S system. In addition, for superconducting PAHs, in the K$_3$picene \cite{ref42}, below 9 THz, the vibration modes come from the motions of potassium as well as the coupling between it and organic molecules, the intermolecular coupling vibrations appear in the range of $9-47$ THz, and the high-frequency C-H stretching modes is localized around 91 THz. From this comparison, there is stronger intermolecular coupling vibrations in C$_6$H$_6$ at pressures than K$_3$picene.

\begin{figure}
\includegraphics[width=\columnwidth]{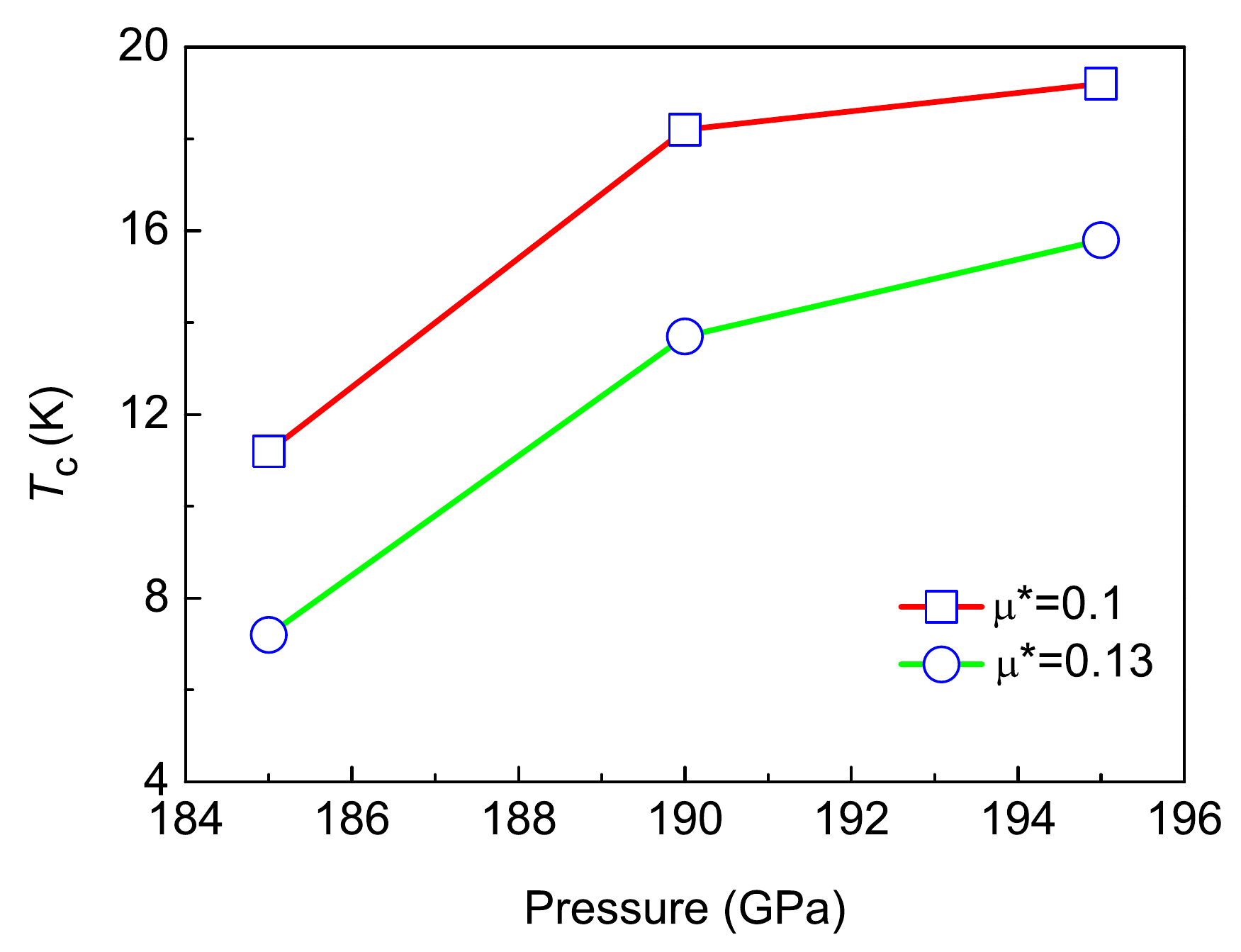}
\caption{(Color online) Pressure dependence of the superconducting critical temperature $T_c$ for $P2_1/c$. The typical value of the Coulomb pseudopotential $\mu^{\star}$ ($0.1-0.13$) is used for calculating $T_c$.}
\end{figure}

Based on electronic and phonon properties above, the electron-phonon coupling $\lambda(\omega)$, logarithmic average phonon frequency $\omega_{log}$ and the Eliashberg phonon spectral function $\alpha^2F(\omega)$ have been investigated to explore the possible superconductivity of C$_6$H$_6$. From the $\alpha^2F(\omega)$ and $\lambda(\omega)$ shown in Fig. 3, the total electron-phonon coupling $\lambda$ is mainly contributed by the low-frequency phonon from C atom, namely the motion of electrons on benzene-ring plane. The rest small account of total $\lambda$ comes from the organic intermolecular coupling. In the case of 190 GPa of $P2_1/c$ structure, the calculated $\lambda$ is 0.68, a moderate-intensity electron-phonon coupling. The low-frequency vibration modes from C atoms contribute 91\% of total $\lambda$, while the organic intermolecular coupling has only 9\% contribution to total $\lambda$, and the high-frequency phonon from C-H bond stretching does not couple electron. Based on the calculated $\alpha^2F(\omega)$, $\lambda$ and $\omega_{log}$, using commonly accepted values of the Coulomb pseudopotential $\mu^{\star}$ ($0.1-0.13$) from the modified McMillan equation by Allen and Dynes \cite{ref43}, however, the superconducting critical temperature of C$_6$H$_6$ was obtained in the range of 13.7 K to 18.2 K at 190 GPa. Interestedly, the $T_c$ gradually increases with the pressure from 180 GPa to 200 GPa, as shown in Fig. 4. At 195 GPa, the $T_c$ of C$_6$H$_6$ reaches 19.2 K as $\mu^{\star}=0.1$. Thus, we theoretically suggest that benzene is superconducting at high pressure, with the $T_c$ close to 20 K.

By analyzing superconductivity, we find that the total $\lambda$ of K$_3$picene \cite{ref42} is mainly due to the low-frequency phonon coupling to electron from potassium, and their $\lambda$ values, 0.68 for C$_6$H$_6$ and 0.73 for K$_3$picene, are comparable in quantity. The difference is that C$_6$H$_6$ has higher $\omega_{log}$ of 565.1 K at 190 GPa, which is almost two times than that of K$_3$C$_{22}$H$_{14}$. Hence, the predicted $T_c$ of C$_6$H$_6$ is almost two times high than that of K$_3$picene \cite{ref42}. Comparing with hydrogen-rich materials, the same is that the pressure induced the metallization, but the difference is that the total $\lambda$ of hydrogen dominate material is mainly contributed by the intermediate-frequency phonon coupling electron. The higher $\lambda$ is often obtained in hydrogen-rich systems. Especially, the H$_3$S results in the $\lambda$ of 2.19 at 200 GPa. Except for the small $\lambda$ value, the $\omega_{log}$ of C$_6$H$_6$ is less than half of H$_3$S. As a result, the $T_c$ of C$_6$H$_6$ is much less than that of H$_3$S. Combining with the comparison of phonon frequency above, we know that the low and intermediate-frequency phonon coupling electron mainly contributes to the superconductivity. On one hand, the phonon softening in this region will be help for the improvement of the $T_c$ in C$_6$H$_6$. On the other hand, the DOS at Fermi level is only 0.013 states/ev/atom for C$_6$H$_6$ at 190 GPa. This value is much less than 0.083 states/ev/atom for PbH$_4$(H$_2$)$_2$ \cite{ref41} and 0.046 states/ev/atom for H$_3$S \cite{ref6}. So the more Cooper-pair electrons is another key to achieve the stronger electron-phonon coupling in C$_6$H$_6$. For instant, pressure leads to the increase of DOS at Fermi level, thus the $T_c$ rises with the increasing pressure in C$_6$H$_6$.

In conclusion, with the aim of exploring superconductivity in a compound containing only carbon and hydrogen, we have studied electronic structures, dynamics properties, and electron-phonon interactions of C$_6$H$_6$ at high pressures. We revealed that C$_6$H$_6$ is superconducting in the pressure range of $180-200$ GPa, which is another evidence of that the system containing benzene rings must be superconducting. $T_c$ was found to gradually increase with the increase of pressure, reaching 20 K at 195 GPa. Differing from the H-dominated materials, the phonon vibrations of C atom mainly contribute to the electron-phonon coupling. The $T_c$ higher than that of K-doped benzene, phenanthrene and picene mainly comes from the larger $\omega_{log}$ induced by pressure. The prediction of superconductivity in C$_6$H$_6$ under pressure will call for experimental testing and the comparison of effects from the electron-phonon coupling and the electronic correlation in such a light element material.

The work was supported by the National Natural Science Foundation of China (Grant Nos. 11274335 and 61574157) and the Basic Research Program of Shenzhen (JCYJ20160331193059332). The partial calculation was supported by the Special Program for Applied Research on Super Computation of the NSFC-Guangdong Joint Fund (the second phase) under Grant No. U1501501.

\bibliography{aipsamp}

\end{document}